\documentclass{PoS}
\pdfoutput=1
\usepackage{amsmath}

\DeclareMathOperator\tr{tr}

\DeclareMathOperator\sign{sgn}
\newcommand\Sp{\text{Sp}}
\newcommand\SU{\text{SU}}
\newcommand\U{\text{U}}
\newcommand\ph{\text{phys}}
\newcommand\hm{\hat m}
\newcommand\hmu{\hat\mu}
\newcommand\ev[1]{\left\langle#1\right\rangle}

\title{Exact results for two-color QCD at low and high
  density\thanks{Supported by the German Research Foundation (DFG) and
    by JSPS.}}

\ShortTitle{Exact results for two-color QCD at low and high density}

\author{Takuya Kanazawa\\
  Department of Physics, The University of Tokyo, Tokyo
  113-0033, Japan\\
  E-mail: \email{tkanazawa@nt.phys.s.u-tokyo.ac.jp}}
\author{\speaker{Tilo Wettig}\\
  Department of Physics, University of Regensburg, 93040
  Regensburg, Germany\\
  E-mail: \email{tilo.wettig@physik.uni-regensburg.de}}
\author{Naoki Yamamoto\\
  Institute for Nuclear Theory, University of Washington,
  Seattle, WA 98195-1550, USA\\ 
  E-mail: \email{nyama@phys.washington.edu}}

\abstract{We discuss a random matrix theory that was originally
  constructed to describe two-color QCD at low density in the phase
  with a nonzero chiral condensate.  With a particular choice of a
  parameter, the same random matrix theory also describes the
  high-density phase of two-color QCD.  In this phase a BCS superfluid
  of diquark pairs is formed, and the pattern of chiral symmetry
  breaking is very different from that at low density.  Analytical
  results for the spectral density obtained from this random matrix
  theory allow for the extraction of the BCS gap from lattice data.}

\FullConference{The XXVIII International Symposium on Lattice Field
  Theory, Lattice2010\\ 
  June 14-19, 2010\\
  Villasimius, Italy}

\begin{document}

\section{Introduction}

Lattice studies of QCD at nonzero quark chemical potential $\mu$ are
hindered by the infamous sign problem, see \cite{deForcrand:2010ys}
for a review.  Two-color QCD with an even number of pairwise
degenerate quarks does not have a sign problem and can therefore be
simulated on the lattice \cite{Kogut:2001na}.  It shares many
qualitative features, such as confinement and chiral symmetry
breaking, with three-color QCD, but the detailed features of both
theories are rather different, such as the pattern of chiral symmetry
breaking, the particle spectrum, or the phase diagram.  Nevertheless,
two-color QCD is an interesting theory in its own right.  It has been
studied in great detail at zero and low density, see, e.g.,
\cite{Kogut:2000ek}.  In this contribution, we also address the region
of high density in which the pattern of chiral symmetry breaking is
different from that at low density and in which a BCS superfluid of
diquark pairs is expected to be formed because there is an attractive
channel between quarks near the Fermi surface.  In earlier work
\cite{Kanazawa:2009ks}, we have derived the low-energy effective
chiral Lagrangian for $\mu\gg\Lambda_{\SU(2)}$, identified the
corresponding $\varepsilon$-regime, and derived Leutwyler-Smilga-type
sum rules for the eigenvalues of the Dirac operator.  This work has
been summarized at Lattice 2009 \cite{Kanazawa:2009pc}.

In the lowest order of the $\varepsilon$-regime, sometimes also called
``microscopic domain'', the theory becomes zero-dimensional.  This
zero-dimen\-sio\-nal limit of the theory can alternatively be
described by a random matrix theory (RMT).  Many examples of such an
exact mapping are known, in particular for two- and three-color QCD at
zero and low density, see \cite{Verbaarschot:2000dy,Akemann:2007rf}
for reviews.  Therefore the natural question is what random matrix
theory describes the microscopic domain of two-color QCD at high
density.  The answer to this question was given in
\cite{Kanazawa:2009en} and will be reviewed in Sec.~\ref{sec:rmt}.
The advantage of having a random matrix theory is that it allows us to
compute a large number of analytical results characterizing the Dirac
eigenvalues, see Sec.~\ref{sec:results}.  This task would be much more
difficult in the effective theory.  The analytical results at high
density contain the BCS gap $\Delta$, which was computed for
asymptotically high density in a weak-coupling approach in
\cite{Son:1998uk,Schafer:2002yy}, as a parameter.  Therefore, $\Delta$
can be extracted from lattice data for the Dirac eigenvalues by
matching them to the analytical results from random matrix theory.
Another interesting feature of two-color QCD is that it allows us to
study the sign problem, either for an odd number $N_f$ of flavors by
turning on $\mu$, or for even $N_f$ by detuning the quark masses from
their degenerate values, see Sec.~\ref{sec:sign}.

\section{Random matrix theory at low and high density}
\label{sec:rmt}

The pertinent random matrix theory for two-color QCD at low density
has been formulated in \cite{Akemann:2008mp}, with partition function
\begin{align}
  \label{eq:rmt1}
  Z_\text{RMT}(\mu)=\int dPdQ\: e^{-\frac12\tr(PP^T+QQ^T)}
  \prod_{f=1}^{N_f}\det \begin{pmatrix}
    m_f & P+\mu Q\\
    -P^T+\mu Q^T & m_f
  \end{pmatrix}\,,
\end{align}
where the $m_f$ are the quark masses, $P$ and $Q$ are real matrices of
dimension $N\times(N+\nu)$, $dP$ and $dQ$ are Cartesian integration
measures, $N$ is assumed to be proportional to the Euclidean
space-time volume $V_4$, and $\nu$ can be identified with the
topological charge.  Note that sometimes other conventions for the
width of the Gaussian distribution of $P$ and $Q$ are used in the
literature.  The RMT Dirac operator $D(\mu)$ is the matrix in
\eqref{eq:rmt1} with the mass set to zero.

It was shown in \cite{Kanazawa:2009en} that in the $N\to\infty$ limit
this RMT partition function is identical to the partition function
obtained from the static (or zero-dimensional) effective Lagrangian
for two-color QCD at low density\footnote{By low density we here mean
  the regime of weak non-Hermiticity, see Sec.~\ref{sec:results} for
  the definition of this regime.} given in \cite{Kogut:2000ek}.  More
precisely, the two partition functions have the same dependence on the
quark masses and on the chemical potential.  The mapping between
dimensionless RMT quantities and physical quantities is given by
\begin{subequations}
  \label{eq:resc1}
  \begin{align}
    \label{eq:resc1a}
    \sqrt Nm&=m_\ph GV_4=m_\pi^2F^2V_4\,,\\
    \frac12N\mu^2&=\mu_\ph^2F^2V_4\,,
  \end{align}
\end{subequations}
where $G$ and $F$ are low-energy constants in the effective Lagrangian
in the notation of \cite{Kogut:2000ek}.  

Both the random matrix theory \eqref{eq:rmt1} and the corresponding
effective Lagrangian explicitly depend on the chemical potential
$\mu$.  In contrast, the effective Lagrangian at high density derived
in \cite{Kanazawa:2009ks} does not explicitly depend on $\mu$.  It
only depends on the quark masses, which appear in the combination
$m^2\Delta^2V_4$.  A hint as to what the correct random matrix theory
at high density should be can be obtained by noting that
\eqref{eq:rmt1} is basically symmetric under $\mu\to1/\mu$ (except
that real and imaginary parts are interchanged).  Maximum
non-Hermiticity, which is expected at high density, corresponds to
$\mu=1$.  We therefore conjecture that at high density the random
matrix theory is, after a redefinition of the random matrices, given
by
\begin{align}
  \label{eq:rmt2}
  Z_\text{RMT}=\int dAdB\: e^{-\frac14\tr(AA^T+BB^T)}
  \prod_{f=1}^{N_f}\det
  \begin{pmatrix}
    m_f & A\\
    B^T & m_f
  \end{pmatrix}\,,
\end{align} 
where the dimension of $A$ and $B$ is again
$N\times(N+\nu)$.\footnote{Only the case $\nu=0$ is physically
  relevant since topology is strongly suppressed at high density.} In
the high-density phase, we restrict ourselves to an even number of
flavors.

Let us first check that we obtain the correct pattern of chiral
symmetry breaking.  To this end, we rewrite the $N_f$-flavor
determinant resulting from \eqref{eq:rmt2} in the chiral limit in the
form
\begin{align}
  {\det}^{N_f}\begin{pmatrix}
    0 & A\\
    B^T & 0
  \end{pmatrix}
  ={\det}^{N_f/2}\begin{pmatrix}
    0 & A\\
    -A^T & 0
  \end{pmatrix}
  {\det}^{N_f/2}\begin{pmatrix}
    0 & B\\
    -B^T & 0
  \end{pmatrix}\,.
\end{align}
The matrices in the two factors on the RHS of this equation have the
form of the chiral orthogonal ensemble of random matrix theory.  It
was shown in \cite{Halasz:1995qb} that the symmetry breaking pattern
in that ensemble with $N_f/2$ flavors is $\U(N_f)\to\Sp(N_f)$.  Since
we have two such factors, \eqref{eq:rmt2} with $N_f$ flavors has the
symmetry breaking pattern
$\U(N_f)\times\U(N_f)\to\Sp(N_f)\times\Sp(N_f)$.  This agrees with the
symmetry breaking pattern in the effective theory due to the formation
of a diquark condensate, which is given by
$\SU(N_f)_L\times\SU(N_f)_R\times\U(1)_B\times\U(1)_A \to
\Sp(N_f)_L\times\Sp(N_f)_R$ \cite{Kanazawa:2009ks}.

We have also shown \cite{Kanazawa:2009en} that in the $N\to\infty$
limit the RMT partition function \eqref{eq:rmt2} is identical to the
partition function of the high-density effective theory in the
zero-dimensional limit, i.e., the two partition functions have the
same mass dependence.  The mapping between the dimensionless RMT mass
and the physical mass is now quite different from \eqref{eq:resc1a}
and given by
\begin{align}
  \label{eq:resc2}
  m=\frac{\sqrt3}{\pi}m_\ph\Delta\sqrt{V_4}\:.
\end{align}

The arguments presented so far, while giving overwhelming evidence in
favor of the equivalence of the random matrix theory \eqref{eq:rmt2}
and the effective theory at high density, do not constitute a full
proof.  For such a proof one would have to show that all spectral
correlation functions are identical in both theories, which requires
studying the partially quenched version of the theory.  Such a study
has not been done yet, but we have no doubt that the outcome would be
positive.

\section{Exact results from random matrix theory}
\label{sec:results}

We can now proceed to compute spectral correlation functions from the
random matrix theory in the $N\to\infty$ limit.  At $\mu=0$ the RMT
eigenvalues $\lambda$ are purely imaginary, while at $\mu\ne0$ they
are either purely real, purely imaginary, or come in complex conjugate
pairs.  We are mainly interested in the so-called microscopic spectral
density of the small eigenvalues, i.e., we rescale all eigenvalues by
a quantity $\delta$ that is, up to a numerical prefactor, equal to the
mean level spacing near zero.  This results in complex numbers
$z=\lambda/\delta$ of order $O(1)$.  To see an effect of the quark
masses on the small eigenvalues we need to rescale them in the same
way, resulting in $\hm_f=m_f/\delta$.

The random matrix theory can be solved in two different regimes:
\begin{itemize}
\item In the regime of weak non-Hermiticity, the combination
  $\hmu^2=2N\mu^2=4\mu_\ph^2F^2V_4$ is kept fixed in the limit
  $N\to\infty$.  While this regime might appear to be mainly of
  academic interest since $\mu\to0$ in the thermodynamic limit, it has
  an important phenomenological application, i.e., the extraction of
  the low-energy constants $G$ and $F$ from lattice data.  In our
  conventions we have $\delta=1/2\sqrt N$ in this regime, i.e.,
  $z=2\sqrt N\lambda$ and $\hm=2\sqrt Nm$.
\item In the regime of strong non-Hermiticity, $\mu$ is kept nonzero
  in the limit $N\to\infty$.  The analytical results in this regime
  are the $\hmu\to\infty$ limits of the corresponding weak
  non-Hermiticity results.  Their functional form is identical for all
  $0<\mu\le1$, and the $\mu$-dependence only enters through a rescaling
  of the eigenvalues.  In our conventions we have $\delta=1$ in this
  regime so that no $N$-dependent rescaling of the eigenvalues and the
  masses is necessary.
\end{itemize}

\begin{figure}
  \centering
  \includegraphics[height=30mm]{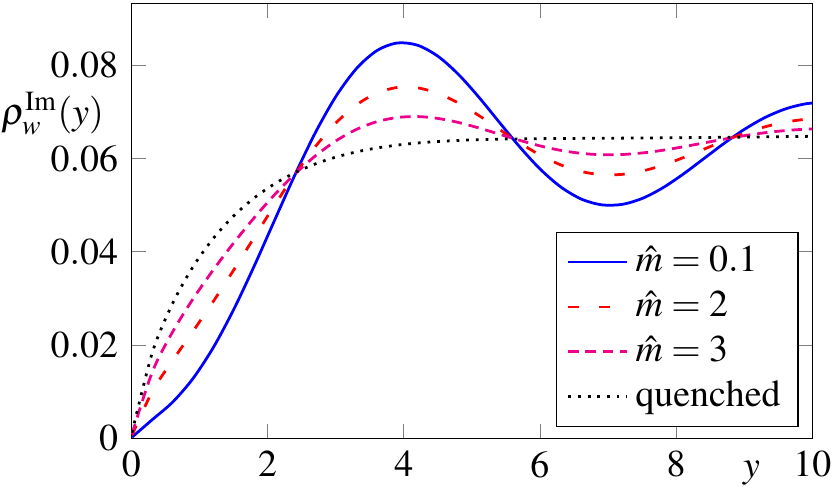}
  \hspace*{20mm}
  \includegraphics[height=30mm]{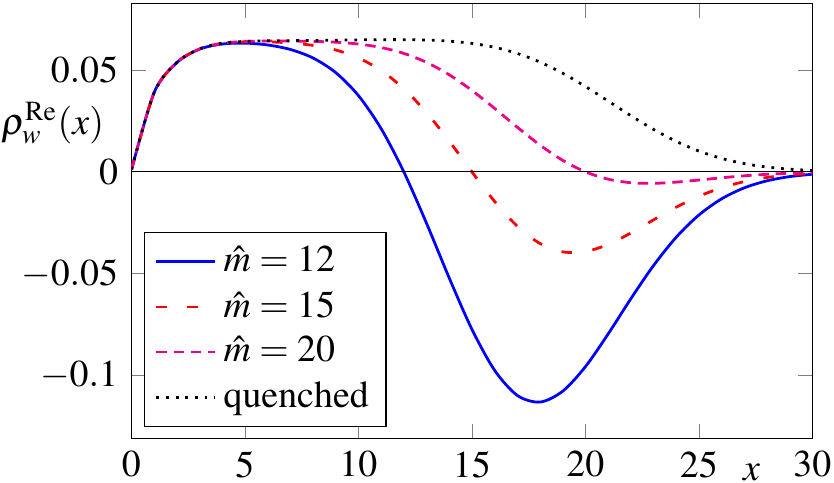}
  \caption{Microscopic spectral ``density'' of the purely imaginary
    (left) and purely real (right) eigenvalues in the regime of weak
    non-Hermiticity for $N_f=1$, $\hmu=3$, $\nu=0$, and different
    values of $\hm$.  The density of the real eigenvalues goes through
    zero for $x=\hm_f$.}
  \label{fig:reim_weak}
\end{figure}

\begin{figure}
  \centering  
  \includegraphics[height=40mm]{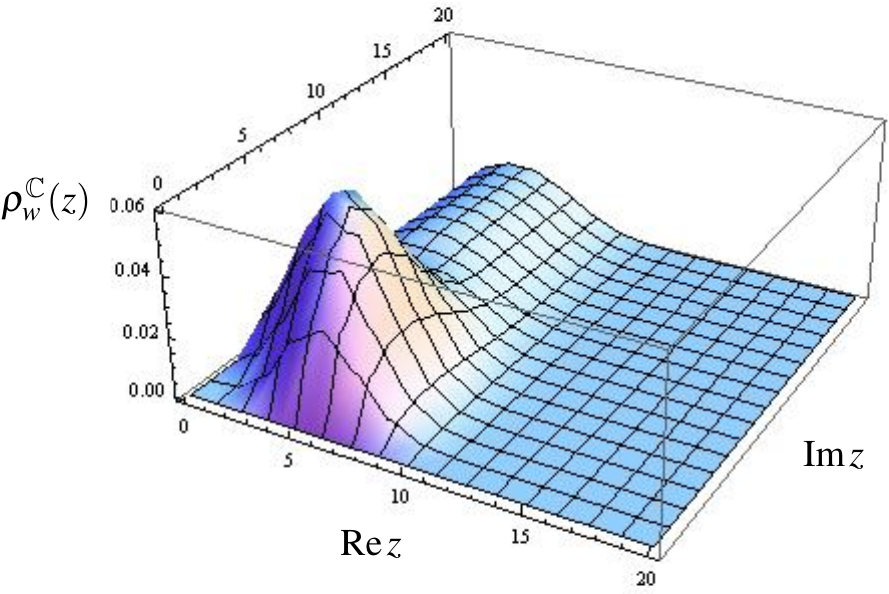}
  \hspace*{20mm}
  \includegraphics[height=40mm]{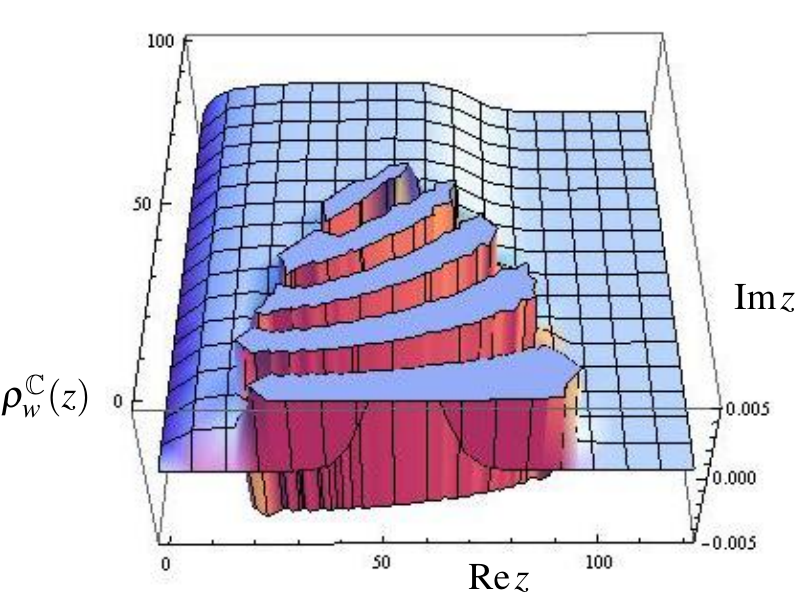}
  \caption{Microscopic spectral ``density'' of the complex eigenvalues
    in the regime of weak non-Hermiticity for $N_f=1$ and $\nu=0$.
    Left: $\hmu=1.8$ and $\hm=0$.  The massless quark causes a
    depletion of the density near the origin.  Right: $\hmu=6$ and
    $\hm=20$.  For large $\hmu$ there is an elliptical domain in which
    the ``density'' oscillates strongly.}
  \label{fig:complex_weak}
\end{figure}

\begin{figure}
  \centering
  \includegraphics[height=30mm]{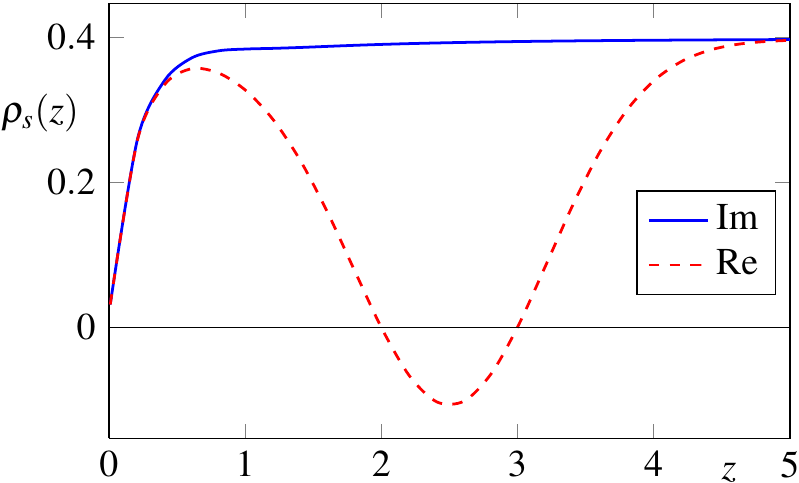}
  \hspace*{20mm}
  \includegraphics[height=30mm]{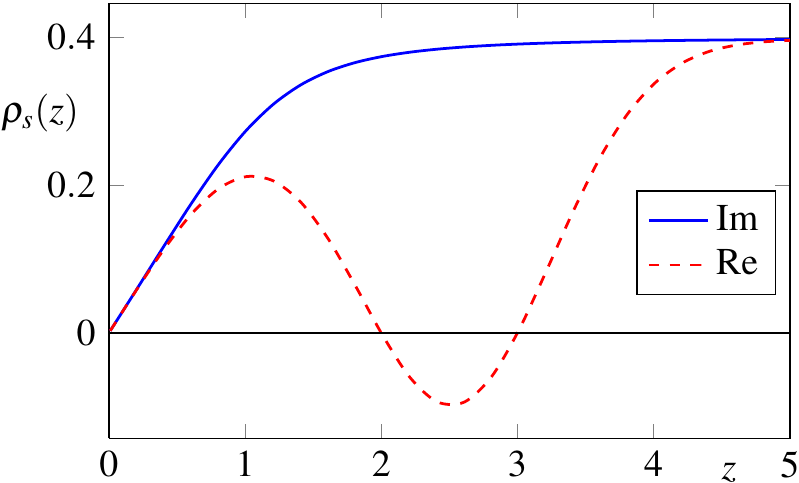}
  \caption{Microscopic spectral ``density'' of the purely imaginary
    (solid) and purely real (dashed) eigenvalues in the regime of
    strong non-Hermiticity for $N_f=2$, $\hm_1=2$, $\hm_2=3$, $\nu=0$
    (left), and $\nu=2$ (right).  The density of the real eigenvalues
    goes through zero for $x=\hm_f$.}
  \label{fig:reim_strong}
\end{figure}

\begin{figure}
  \centering
  \includegraphics[height=40mm]{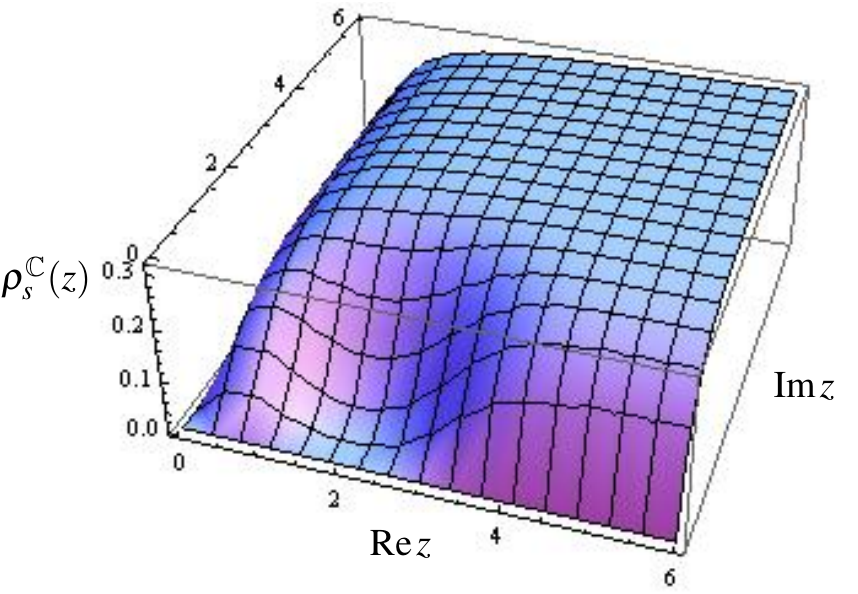}
  \hspace*{20mm}
  \includegraphics[height=40mm]{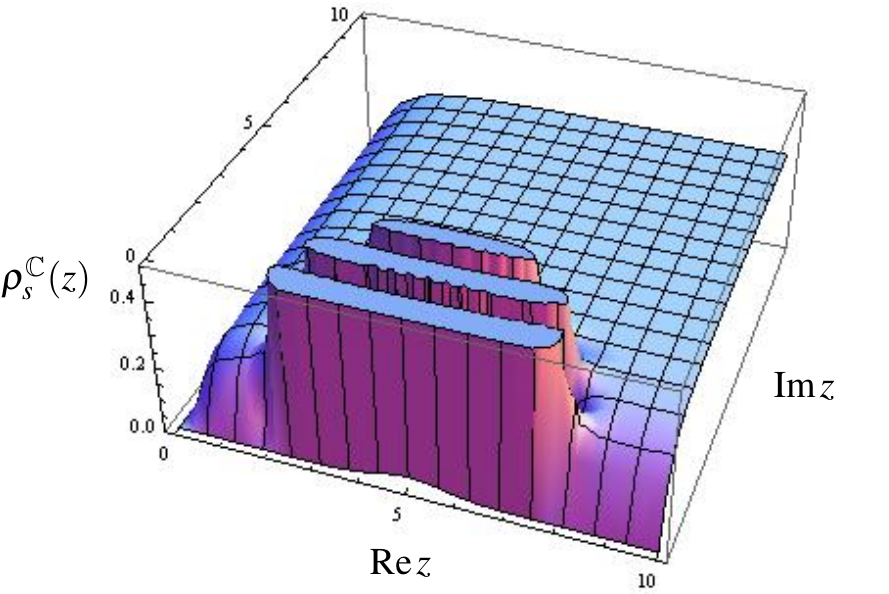}
  \caption{Microscopic spectral ``density'' of the complex eigenvalues
    in the regime of strong non-Hermiticity for $N_f=2$ and $\nu=0$.
    Left: $\hm_1=\hm_2=2$.  Degenerate masses generate a dip in the
    spectrum at $z=\hm$. Right: $\hm_1=2$, $\hm_2=8$.  Unequal masses
    result in a domain of strong oscillations, indicative of the sign
    problem.}
  \label{fig:complex_strong}
\end{figure}

In \cite{Akemann:2009fc} analytical results for the microscopic
spectral ``density'' (which we put in quotation marks since this
quantity can become negative if there is a sign problem) were
obtained in both of the above-mentioned regimes in the quenched case,
i.e., for $N_f=0$ flavors.  In the meantime, the generalization to the
unquenched case has been worked out \cite{Akemann:2010tbd}.  Since the
analytical results are rather cumbersome we will not present them
here.  Important features of the results are exhibited in
Figs.~\ref{fig:reim_weak} through \ref{fig:complex_strong}.  Comments
on these features are given in the figure captions.

\section{The sign problem}
\label{sec:sign}

As a good measure of the sign problem in two-color QCD we define the
quantity
\begin{align}
  \ev{\sign\:\det(D+m)}_{||N_f||}\equiv
  \frac{\ev{\sign\:\det(D+m)\prod_{f=1}^{N_f}|\det(D+m_f)|}_{N_f=0}}
  {\ev{\prod_{f=1}^{N_f}|\det(D+m_f)|}_{N_f=0}}\,,
\end{align}
see \cite{Akemann:2010tbd} for a more detailed discussion.  

\begin{figure}
  \centering
  \includegraphics[scale=.7]{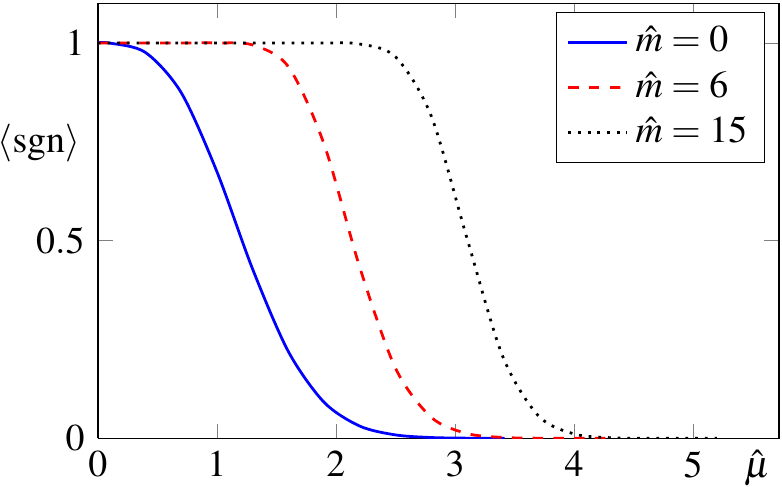}
  \hspace*{20mm}
  \includegraphics[scale=.7]{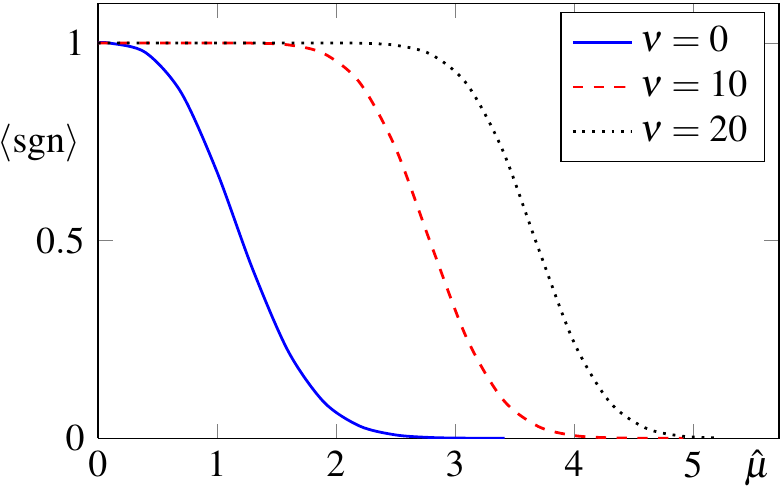}
  \caption{Average sign at weak non-Hermiticity for $N_f=1$ as a
    function of $\hmu$.  Left: $\nu=0$ is kept fixed and $\hm$ is
    varied.  Right: $\hm=0$ is kept fixed and $\nu$ is varied.}
  \label{fig:sign_weak}
\end{figure}
We first consider the regime of weak non-Hermiticity.  We choose
$N_f=1$ and turn on $\hmu$ to study its effect on the average sign,
see Fig.~\ref{fig:sign_weak}.  It is evident from the plots that the
sign problem (i) increases with increasing $\hmu$, (ii) decreases with
increasing $\hm$, and (iii) decreases with increasing $\nu$ (in
agreement with \cite{Bloch:2008cf}).  A quantitative analysis
\cite{Akemann:2010tbd} reveals that in the thermodynamic limit the
average sign makes a first-order transition from $1$ to $0$ at
$\hmu=\sqrt{\hm/2}$, which in physical units corresponds to a critical
chemical potential $\mu_\ph=m_\pi/2$.

\begin{figure}
  \centering
  \includegraphics[scale=.7]{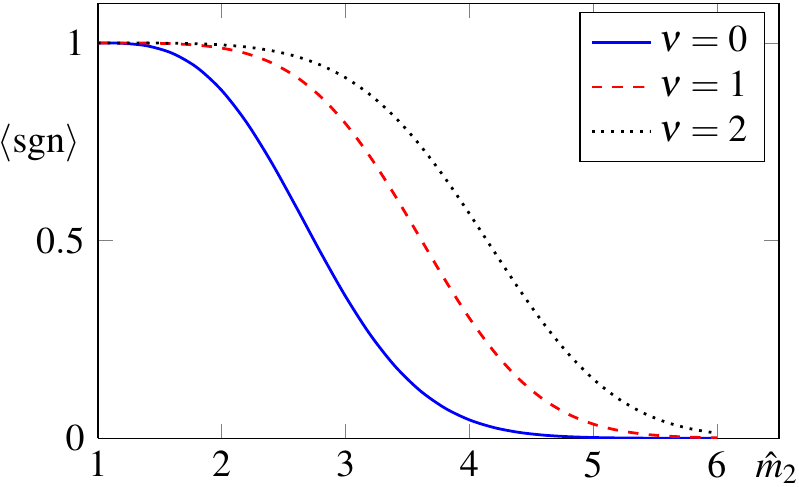}
  \caption{Average sign at strong non-Hermiticity for $N_f=2$,
    $\hm_1=1$, and $\nu=0,1,2$ as a function of $\hm_2$.}
  \label{fig:sign_strong}
\end{figure}
Next we consider the regime of strong non-Hermiticity.  We choose
$N_f=2$ and detune the quark masses.  The effect on the average sign
is shown in Fig.~\ref{fig:sign_strong}.  The sign problem is absent
for $\hm_1=\hm_2$ and increases as $|\hm_1-\hm_2|$ increases.  It
again decreases with increasing $\nu$.

\section{Conclusions}

We have shown that a single random matrix theory describes two-color
QCD at low density in the regime of weak-Hermiticity and at high
density in the BCS superfluid phase, depending on the choice of the
RMT parameter $\mu$ and on the rescaling factors in \eqref{eq:resc1}
and \eqref{eq:resc2}.  These two regimes have very different symmetry
breaking patterns.  It would be interesting to investigate the
applicability of random matrix theory in the region of intermediate
densities, where intriguing phenomena such as a BEC-BCS crossover have
been conjectured \cite{Splittorff:2000mm}.

The analytical RMT results can be used to extract physical parameters
such as $\Delta$ from lattice data.  Two-color lattice simulations
with adjoint staggered fermions, which are in the same symmetry class
as continuum fundamental fermions \cite{Hands:2000ei}, are currently
underway to test our RMT predictions.

\bibliographystyle{JHEP}
\bibliography{tilo}

\providecommand{\href}[2]{#2}\begingroup\raggedright\begin{thebibliography}{10}

\bibitem{deForcrand:2010ys}
P.~de~Forcrand, {\it {Simulating QCD at finite density}},  {\em PoS} {\bf
  LAT2009} (2009) 010, [\href{http://arxiv.org/abs/1005.0539}{{\tt
  arXiv:1005.0539}}].

\bibitem{Kogut:2001na}
J.~B. Kogut, D.~K. Sinclair, S.~J. Hands, and S.~E. Morrison, {\it {Two-colour
  QCD at non-zero quark-number density}},  {\em Phys. Rev.} {\bf D64} (2001)
  094505, [\href{http://arxiv.org/abs/hep-lat/0105026}{{\tt hep-lat/0105026}}].

\bibitem{Kogut:2000ek}
J.~B. Kogut, M.~A. Stephanov, D.~Toublan, J.~J.~M. Verbaarschot, and
  A.~Zhitnitsky, {\it {QCD-like theories at finite baryon density}},  {\em
  Nucl. Phys.} {\bf B582} (2000) 477--513,
  [\href{http://arxiv.org/abs/hep-ph/0001171}{{\tt hep-ph/0001171}}].

\bibitem{Kanazawa:2009ks}
T.~Kanazawa, T.~Wettig, and N.~Yamamoto, {\it {Chiral Lagrangian and spectral
  sum rules for dense two-color QCD}},  {\em JHEP} {\bf 08} (2009) 003,
  [\href{http://arxiv.org/abs/0906.3579}{{\tt arXiv:0906.3579}}].

\bibitem{Kanazawa:2009pc}
T.~Kanazawa, T.~Wettig, and N.~Yamamoto, {\it {Chiral Lagrangian and spectral
  sum rules for two-color QCD at high density}},  {\em PoS} {\bf LAT2009}
  (2009) 195, [\href{http://arxiv.org/abs/0910.2300}{{\tt arXiv:0910.2300}}].

\bibitem{Verbaarschot:2000dy}
J.~J.~M. Verbaarschot and T.~Wettig, {\it {Random matrix theory and chiral
  symmetry in QCD}},  {\em Ann. Rev. Nucl. Part. Sci.} {\bf 50} (2000)
  343--410, [\href{http://arxiv.org/abs/hep-ph/0003017}{{\tt hep-ph/0003017}}].

\bibitem{Akemann:2007rf}
G.~Akemann, {\it {Matrix models and QCD with chemical potential}},  {\em Int.
  J. Mod. Phys.} {\bf A22} (2007) 1077--1122,
  [\href{http://arxiv.org/abs/hep-th/0701175}{{\tt hep-th/0701175}}].

\bibitem{Kanazawa:2009en}
T.~Kanazawa, T.~Wettig, and N.~Yamamoto, {\it {Chiral random matrix theory for
  two-color QCD at high density}},  {\em Phys. Rev.} {\bf D81} (2010) 081701,
  [\href{http://arxiv.org/abs/0912.4999}{{\tt arXiv:0912.4999}}].

\bibitem{Son:1998uk}
D.~T. Son, {\it {Superconductivity by long-range color magnetic interaction in
  high-density quark matter}},  {\em Phys. Rev.} {\bf D59} (1999) 094019,
  [\href{http://arxiv.org/abs/hep-ph/9812287}{{\tt hep-ph/9812287}}].

\bibitem{Schafer:2002yy}
T.~Schafer, {\it {QCD and the eta' mass: Instantons or confinement?}},  {\em
  Phys. Rev.} {\bf D67} (2003) 074502,
  [\href{http://arxiv.org/abs/hep-lat/0211035}{{\tt hep-lat/0211035}}].

\bibitem{Akemann:2008mp}
G.~Akemann, M.~J. Phillips, and H.~J. Sommers, {\it {Characteristic polynomials
  in real Ginibre ensembles}},  {\em J. Phys.} {\bf A42} (2008) 012001,
  [\href{http://arxiv.org/abs/0810.1458}{{\tt arXiv:0810.1458}}].

\bibitem{Halasz:1995qb}
A.~M. Halasz and J.~J.~M. Verbaarschot, {\it {Effective Lagrangians and chiral
  random matrix theory}},  {\em Phys. Rev.} {\bf D52} (1995) 2563--2573,
  [\href{http://arxiv.org/abs/hep-th/9502096}{{\tt hep-th/9502096}}].

\bibitem{Akemann:2009fc}
G.~Akemann, M.~J. Phillips, and H.~J. Sommers, {\it {The chiral Gaussian
  two-matrix ensemble of real asymmetric matrices}},  {\em J. Phys.} {\bf A43}
  (2010) 085211, [\href{http://arxiv.org/abs/0911.1276}{{\tt
  arXiv:0911.1276}}].

\bibitem{Akemann:2010tbd}
G.~Akemann, T.~Kanazawa, M.~Phillips, and T.~Wettig, {\it {Random matrix theory
  of unquenched two-colour QCD with nonzero chemical potential}},
[\href{http://arxiv.org/abs/1012.4461}{{\tt arXiv:1012.4461}}].

\bibitem{Bloch:2008cf}
J.~C.~R. Bloch and T.~Wettig, {\it {Random matrix analysis of the QCD sign
  problem for general topology}},  {\em JHEP} {\bf 03} (2009) 100,
  [\href{http://arxiv.org/abs/0812.0324}{{\tt arXiv:0812.0324}}].

\bibitem{Splittorff:2000mm}
K.~Splittorff, D.~T. Son, and M.~A. Stephanov, {\it {QCD-like Theories at
  Finite Baryon and Isospin Density}},  {\em Phys. Rev.} {\bf D64} (2001)
  016003, [\href{http://arxiv.org/abs/hep-ph/0012274}{{\tt hep-ph/0012274}}].

\bibitem{Hands:2000ei}
S.~Hands {\em et~al.}, {\it {Numerical study of dense adjoint matter in two
  color QCD}},  {\em Eur. Phys. J.} {\bf C17} (2000) 285--302,
  [\href{http://arxiv.org/abs/hep-lat/0006018}{{\tt hep-lat/0006018}}].

\end{thebibliography}\endgroup

\end{document}